\documentclass[preprint2]{emulateapj}
\usepackage{graphicx}
\usepackage{txfonts}
\usepackage{url}
\graphicspath{{./fig/}{./png/}}
\newcommand{\EQ}{\begin{equation}}
\newcommand{\EN}{\end{equation}}
\newcommand{\EQA}{\begin{eqnarray}}
\newcommand{\ENA}{\end{eqnarray}}

\newcommand{\Eq}[1]{equation~(\ref{#1})}

\newcommand{\Fig}[1]{Fig.~\ref{#1}}

\newcommand{\Tab}[1]{Table~\ref{#1}}

\newcommand{\bra}[1]{\langle #1\rangle}

\newcommand{\meanrho}{\overline{\rho}}

{}
{}
\newcommand{\meanemfs}{\overline{\cal E} {}}

{}
{}
{}
\newcommand{\meanEMF}{\overline{\mbox{\boldmath ${\cal E}$}}{}}{}
{}
{}
{}
{}
{}
\newcommand{\meanBB}{\overline{\mbox{\boldmath $B$}}{}}{}
\newcommand{\meanJJ}{\overline{\mbox{\boldmath $J$}}{}}{}
\newcommand{\meanUU}{\overline{\mbox{\boldmath $U$}}{}}{}
{}
{}

\newcommand{\meanB}{\overline{B}}

\newcommand{\meanJ}{\overline{J}}

{}
{}
{}
{}
{}

\newcommand{\meanBhat}{\hat{B}}

\newcommand{\uu}{\mbox{\boldmath $u$} {}}
\newcommand{\UU}{\mbox{\boldmath $U$} {}}
\newcommand{\xx}{\mbox{\boldmath $x$} {}}
\newcommand{\aaa}{\mbox{\boldmath $a$} {}}
\newcommand{\bb}{\mbox{\boldmath $b$} {}}
\newcommand{\BB}{\mbox{\boldmath $B$} {}}

\newcommand{\jj}{\mbox{\boldmath $j$} {}}
\newcommand{\JJ}{\mbox{\boldmath $J$} {}}

\newcommand{\AAA}{\mbox{\boldmath $A$} {}}

\newcommand{\ff}{\mbox{\boldmath $f$} {}}

\newcommand{\nab}{\mbox{\boldmath $\nabla$} {}}

\newcommand{\oo}{\mbox{\boldmath $\omega$} {}}

\newcommand{\SSSS}{\mbox{\boldmath ${\sf S}$} {}}

\def\Pm{\mbox{\rm Pr}_M}
\def\Rm{\mbox{\rm Re}_M}

\def\onethird{{\textstyle{1\over3}}}

\newcommand{\yapj}[3]{ #1, {ApJ,} {#2}, #3}

\newcommand{\yapjl}[3]{ #1, {ApJ,} {#2}, #3}

\newcommand{\yan}[3]{ #1, {Astron.\ Nachr.,} {#2}, #3}

\newcommand{\yana}[3]{ #1, {A\&A,} {#2}, #3}

\newcommand{\yjfm}[3]{ #1, {J.\ Fluid Mech.,} {#2}, #3}

\newcommand{\ymn}[3]{ #1, {MNRAS,} {#2}, #3}

\newcommand{\ypre}[3]{ #1, {Phys.\ Rev.\ E,} {#2}, #3}

\newcommand{\yjour}[4]{ #1, {#2}, {#3}, #4}

{}
\newcommand{\alptrel}{\tilde\alpha}
\newcommand{\etatrel}{\tilde\eta_{\it t}}
\newcommand{\etat}{\eta_{\it t}}

\newcommand{\alphaK}{\alpha_{\it K}}
\newcommand{\alphaM}{\alpha_{\it M}}
\newcommand{\kmS}{k_{\it m}}
\newcommand{\kkf}{\mbox{\boldmath $k$}_{\it f}}
\newcommand{\kf}{k_{\it f}}
\newcommand{\cs}{c_{\it s}}

\submitted{Received May 12, 2008, accepted September 12, 2008}
\begin{document}

\title{Magnetic quenching of alpha and diffusivity tensors in helical turbulence}
\author{Axel Brandenburg\altaffilmark{1}, Karl-Heinz R\"adler\altaffilmark{2},
Matthias Rheinhardt\altaffilmark{2}, and Kandaswamy Subramanian\altaffilmark{3}}

\altaffiltext{1}{
NORDITA, Roslagstullsbacken 23, SE-10691 Stockholm, Sweden
}\altaffiltext{2}{
Astrophysical Institute Potsdam, An der Sternwarte 16, D-14482 Potsdam, Germany
}\altaffiltext{3}{
Inter University Centre for Astronomy and Astrophysics,
Post Bag 4, Pune University Campus, Ganeshkhind, Pune 411 007, India
\\ $ $Revision: 1.125 $ $ (\today)
}

\begin{abstract}
The effect of a dynamo-generated mean magnetic field of Beltrami type
on the mean electromotive force is studied.
In the absence of the mean magnetic field
the turbulence is assumed to be homogeneous and isotropic,
but it becomes inhomogeneous and anisotropic with this field.
Using the testfield method the dependence of
the alpha and turbulent diffusivity tensors
on the magnetic Reynolds number $\Rm$ is determined for magnetic fields
that have reached approximate equipartition with the velocity field.
The tensor components are characterized by a pseudoscalar $\alpha$ and a
scalar turbulent magnetic diffusivity $\etat$.
Increasing $\Rm$ from 2
to 600 reduces $\etat$ by a factor $\approx5$, suggesting that the
quenching of $\etat$ is, in contrast to the 2-dimensional case,
only weakly dependent on $\Rm$. Over the same range of $\Rm$, however,
$\alpha$ is reduced by a factor $\approx14$, which can qualitatively
be explained by a corresponding increase of a magnetic contribution to the
$\alpha$ effect with opposite sign.
The level of fluctuations of $\alpha$ and $\etat$ is only 10\% and 20\% of the
respective kinematic reference values.
\end{abstract}


\section{Introduction}

Magnetic fields in stars and galaxies tend to display large scale spatial
order, and in the case of the Sun also long term temporal order
(the 22 year cycle).
The underlying process
is generally believed to be a turbulent large-scale or mean-field dynamo
-- the simplest of which is an $\alpha^2$ dynamo,
which works with helical turbulence and no mean flows.
This can be modeled by direct numerical simulations in a
periodic box where the flow is driven by helical isotropic forcing.
Corresponding simulations by Brandenburg (2001) show that in the
nonlinear regime there is a resistively slow saturation phase associated with nearly perfect
conservation of magnetic helicity.
This slow saturation imposes tight constraints on the
quenching of the electromotive force.
By comparing with suitable mean field models one can only constrain the
quenching of the full electromotive force, but not the individual
quenchings of $\alpha$ and $\etat$, because
the saturated mean magnetic field of an $\alpha^2$ dynamo tends to become force-free,
so the mean magnetic field and the mean current
density are aligned (Blackman \& Brandenburg 2002; hereafter BB02).
As a consequence
an infinitude of combinations of quenching expressions for $\alpha$ and $\etat$
describe the same saturation behavior.

The saturation of the mean magnetic field is well described by a
mutual cancellation of kinetic and magnetic alpha effects, where the
latter depends on the production rate of mean magnetic helicity.
To reproduce the resistively slow saturation,
both kinetic alpha effect, $\alphaK$, and turbulent magnetic diffusivity,
$\etat$, could be assumed completely unquenched.
This is however an unrealistic simplification (Kleeorin \& Rogachevskii 1999).
Some level of quenching
of $\etat$ was found to be necessary to reproduce the simulations (BB02).

Since the early work of Vainshtein \& Cattaneo (1992),
a lot of effort has gone into determining the quenching of $\alpha$.
It is now clear that
for mean fields defined as volume averages over a periodic box
$\alpha$ is ``catastrophically'' quenched like $\Rm^{-1}$ with mean fields
of equipartition strength (Cattaneo \& Hughes 1996).
However, subsequent work showed that this is a particular consequence
of the use of full volume averages, in which case the mean current density
is zero (BB02).

The quenching of $\etat$ is much less understood.
While in the two-dimensional case, $\etat$
is indeed catastrophically quenched (Cattaneo \& Vainshtein 1991),
in three dimensions the quenching may depend just on $\meanBB^2$,
but not on $\Rm$.
This has already been found from the decay rate of a nonhelical large-scale
magnetic field in driven non-helical turbulence (Yousef et al.\ 2003).
Similar indications come also from fitting mean field models to
corresponding simulations (BB02).

Quantifying more precisely the simultaneous quenching of
$\alpha$ and $\etat$ is the goal of the present paper.
We admit both $\alpha$ and $\eta_{\rm t}$
to be tensors, denoted by $\alpha_{ij}$ and $\eta_{ij}$, respectively,
and we calculate them using the testfield method
(e.g., Brandenburg et al.\ 2008, Sur et al.\ 2008).
However, unlike earlier kinematic work, we now allow the
velocity to be the result of the fully nonlinear hydromagnetic equations,
i.e.\ to be influenced by the resulting mean magnetic field.

\section{The method}

Following earlier work by Brandenburg (2001), we consider a compressible
isothermal gas
with sound speed $\cs$,
but in addition we also solve a set of testfield equations, as was done
in Brandenburg et al.\ (2008) for the kinematic case.
The full set of governing equations is then
\EQ
{\partial\UU\over\partial t}=-\UU\cdot\nab\UU-\cs^2\nab\ln\rho
+\ff+\rho^{-1}\Big(\JJ\times\BB+\nab\cdot2\rho\nu\SSSS\Big) \, ,
\label{dUU}
\EN
\EQ
{\partial\ln\rho\over\partial t}=-\UU\cdot\nab\ln\rho-\nab\cdot\UU,
\label{dlnrho}
\EN
\EQ
{\partial\AAA\over\partial t}=\UU\times\BB-\mu_0\eta\JJ, \label{indEq}
\EN
\EQ
{\partial\aaa^{pq}\over\partial t}=\meanUU\times\bb^{pq}+\uu\times\meanBB^{pq}
+\uu\times\bb^{pq}-\overline{\uu\times\bb^{pq}}
-\mu_0\eta\jj^{pq},
\label{nonSOCA}
\EN
where mean fields are defined as horizontal ($xy$) averages,
thus being functions of $z$ and $t$ only, and indicated by overbars
whereas lower case vectors denote deviations from the averages (``fluctuations").
The superscripts $pq$ refer to four separate equations
that are characterized by four different testfields $\meanBB^{pq}$
having a $\cos kz$ or $\sin kz$ dependence ($q={\rm c,s}$)
in the $x$ or $y$ component ($p=1,2$).
We employ a magnetic vector potential both for the
magnetic field $\BB=\nab\times\AAA$ and for the responses to
the testfields, $\bb^{pq}=\nab\times\aaa^{pq}$.
We reinitialize $\aaa^{pq}$ to zero every 30--60 turnover times
to suppress small-scale dynamo action (cf.\ Sur et al.\ 2008).
Of course, the velocity $\UU$ is now affected by the magnetic
field $\BB$ through the Lorentz force.
The current density is $\JJ = \nabla \times \BB/\mu_0$,
where $\mu_0$ is the magnetic permeability.
The flow is driven by random forcing described by a
forcing function $\ff$ consisting of circularly
polarized plane waves with positive helicity and random direction
(giving rise to a flow with maximal helicity), and
${\sf S}_{ij}={1\over2}(U_{i,j}+U_{j,i})-{1\over3}\delta_{ij}\nab\cdot\UU$
is the traceless rate of strain tensor.
The forcing function is chosen such that the moduli of the wavevectors,
$|\kkf|$, are in a narrow interval around an average value,
which is denoted simply by $\kf$.

Owing to our definition of averages, $\meanBB$ is independent of $x$ and $y$
and all its first--order spatial derivatives can be expressed by
the components of $\meanJJ$.
If we ignore higher-order derivatives of $\meanBB$
the mean electromotive force $\meanEMF = \overline{\uu \times \bb}$
has the form
\EQ
\meanemfs_i = \alpha_{ij} \meanB_j - \mu_0\eta_{ij} \meanJ_j \label{meanemf}
\EN
with two tensors $\alpha_{ij}$ and $\eta_{ij}$,
and we restrict our attention to $1 \leq i, j \leq 2$.
For details see Brandenburg et al.\ (2008).
Solving the test field equations allows us to calculate
$\meanEMF^{pq}=\overline{\uu\times\bb^{pq}}$ and, via Eq.~(\ref{meanemf}),
all 4+4 components of $\alpha_{ij}$ and $\eta_{ij}$.

Important control parameters are the magnetic Reynolds and Prandtl numbers,
$\Rm=u_{\rm rms}/(\eta\kf)$ and $\Pm=\nu/\eta$,
where $u_{\rm rms}=\bra{\uu^2}^{1/2}$ is the actual (magnetically
affected) rms velocity and angular brackets denote volume averages.
The smallest possible wavenumber in a triply-periodic domain of size
$L\times L\times L$ is $k_1=2\pi/L$.
In order to achieve large values of $\Rm$, the value of $\kf/k_1$
should be small, but still large enough to allow for a
clear separation of scales between the domain scale and
the energy-carrying scale.
We use $\kf/k_1=3$ as a compromise.

The structure of the turbulence is determined
by the vectors $\meanBB$ and $\meanJJ$,
but for a Beltrami field they are aligned, so we have
\EQ
\alpha_{ij}(\meanBB)=\alpha_1(\meanBB)\delta_{ij}
+\alpha_2(\meanBB)\meanBhat_i\meanBhat_j, \label{alpten}\EN
\EQ
\eta_{ij}(\meanBB)=\eta_1(\meanBB)\delta_{ij}
+\eta_2(\meanBB)\meanBhat_i\meanBhat_j,
\label{etaten}
\EN
where $\hat{\BB}$ means the unit vector in the direction of $\meanBB$.
When inserting this into the general expression for the electromotive
force given above this reduces to $\meanEMF=\alpha\meanBB-\mu_0\etat\meanJJ$, with
coefficients
\EQ
\alpha=\alpha_1+\alpha_2-\eta_2\kmS\quad\mbox{and}\quad
\etat=\eta_1, \label{alpha_etat}
\EN
where $\kmS=\kmS(z,t)\equiv\mu_0\meanJJ\cdot\meanBB/\meanBB^2$ is a
pseudoscalar that quantifies the helicity of the large-scale
field.
(Here $\kmS/k_1\approx-1$.)
We emphasize that for Beltrami fields the assignment of $\alpha_1$, $\alpha_2$,
$\eta_1$ and $\eta_2$ to $\alpha$ and $\eta_{\rm t}$ is not unique.
In the general situation, when the mean field is not of Beltrami type,
instead of $\alpha_2$ and $\eta_2$ eight new
coefficients emerge which contribute in an unambiguous way to field generation and dissipation.
Future work must show whether
our $\alpha_1$ and $\etat$ are then still dominant.

\section{Results}

\begin{figure}[t!]
\centering\includegraphics[width=\columnwidth]{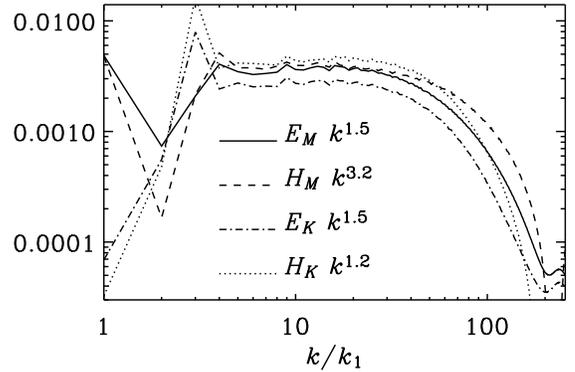}\caption{
Compensated time-averaged spectra of kinetic and magnetic energy, as well as of
kinetic and magnetic helicity, for a run with $\Rm=600$.
}\label{pspec}\end{figure}

\begin{figure}[t!]\begin{center}
\includegraphics[width=\columnwidth]{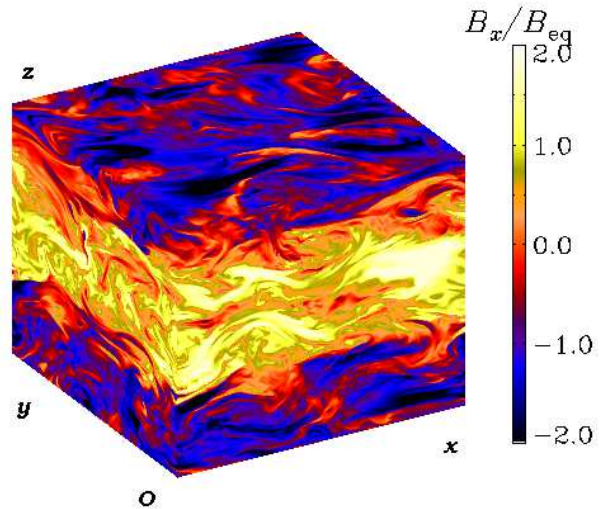}
\end{center}\caption[]{
Visualization of $B_x$ on the periphery of the computational
domain for a run with
$\Rm=600$ and a resolution of $512^3$ mesh points.
Note that on average the field is compatible with that in \Eq{Beltrami}.
Note also the clear anisotropy with structures elongated in the
direction of the field.
For an animation see
\url{http://www.nordita.org/software/pencil-code/movies/icascade/}.
}\label{img_0138}
\end{figure}

Throughout this paper we fix $\Pm=1$ and vary $\Rm$ between 2 and 600.
For large values of $\Rm$ a broader range of scales is excited,
as can be seen in spectra of kinetic and magnetic energy, $E_K$ and $E_M$,
shown in \Fig{pspec}.
In the range $4<k/k_1<30$ both spectra are comparable to a $k^{-3/2}$ spectrum.
For comparison, spectra of kinetic and magnetic helicity, $H_K$ and $H_M$,
are also shown.

For $\Rm\le2$ there is no dynamo action, but in all other cases a large
scale magnetic field is maintained (\Fig{img_0138}), just as in Brandenburg (2001),
except that here $\kf/k_1=3$ instead of 5 or larger.
The dynamo is of $\alpha^2$ type and hence the mean field a Beltrami field,
\EQ
\meanBB(z,t)=\meanB(t)\,(\cos\theta, \sin\theta, 0),\quad
\quad\theta=k_1z+\phi,
\label{Beltrami}
\EN
with phase $\phi$.
To shorten the transient phase we use this field also as initial condition.

Inserting (\ref{Beltrami}) into (\ref{alpten}) and (\ref{etaten}) and calculating suitable averages over $z$ (or volume)
we get
\EQ
\alpha_2=8\left\langle\alpha_{12}\cos\theta\sin\theta\right\rangle =8 \left\langle\alpha_{21}\cos\theta\sin\theta\right\rangle\; ,
\EN
\EQ
 \alpha_1+\alpha_2/2 = \left\langle\alpha_{11}\right\rangle = \left\langle\alpha_{22}\right\rangle\; ,
\EN
and analogous for $\eta_1$ and $\eta_2$.
Obviously, the determination of $\alpha_1$, $\alpha_2$, $\eta_1$, and $\eta_2$
requires knowledge of the Beltrami phase $\phi$,
which often drifts away from its initial value during the course of the run.
We determined therefore
the actual phase $\phi(t)$ by applying a suitable Fourier analysis to $\meanBB$.

In general, $\alpha$ quenching can involve time derivatives
(e.g., Kleeorin \& Ruzmaikin 1982, BB02).
In order to avoid such complications we focus on statistically steady
(dynamo) solutions, that is, on the saturated dynamo fields. For given values
of the parameters of the system (\ref{dUU})--(\ref{indEq}), the saturation strength of
$\meanBB$ is uniquely determined. Hence, by changing the forcing strength or $\eta$ we are
only able to follow a specific path in the $\meanB$ -- $\Rm$ plane, but not to scan it in a 2D fashion.

In \Tab{Tsum} we represent the results in nondimensional form
with normalized quantities indicated by a tilde.
We normalize the rms values of the mean field and the fluctuations with the equipartition field strength
$B_{\rm eq}= (\mu_0\langle\rho u^2(\meanBB)\rangle)^{1/2}$ and introduce
\EQ
\tilde\eta_1=\eta_1/\eta_{t0},\quad
\tilde\eta_2=\eta_2/\eta_{t0},\quad
\tilde\eta=\eta/\eta_{t0},
\label{etatilde}
\EN
\EQ
\tilde\alpha_1=\alpha_1/\alpha_0,\quad
\tilde\alpha_2=\alpha_2/\alpha_0,\quad
\label{alptilde}
\EN
where $\eta_{t0}=\onethird u_{\rm rms}(\meanBB)\kf^{-1}$,
$\alpha_0=-\onethird u_{\rm rms}(\meanBB)$, and
$u_{\rm rms}(\meanBB)$ is the rms velocity of the
saturated state, so the reference values are already magnetically affected.
This normalization implies that in the kinematic case
$\tilde\alpha_1\,=\,\tilde\eta_1\,=\,1$ (Sur et al.\ 2008),
while $\tilde\alpha_2\,=\,\tilde\eta_2=0$.
Error bars are calculated based on the maximum
departure obtained from the three time series,
each taken over one third of the full sequence.

\begin{deluxetable*}{crrrrrlrrrrrrrr} 
  \tabletypesize{\scriptsize} 
  \tablecaption{Transport coefficients for runs
  in the range $2\leq\Rm\leq600$ at saturation field strengths.
  \label{Tsum}
  }
  \tablewidth{0pt}
  \tablehead{
\colhead{Run}&
\colhead{$\Rm$}&\colhead{$\tilde{B}^2$}&\colhead{$\tilde{b}^2$}&
\colhead{$\alptrel$}& \colhead{$\etatrel$}&
\colhead{$\tilde\eta$}&
\colhead{$\tilde\lambda$}&
\colhead{$-\tilde\alpha_2$}& \colhead{$-\tilde\eta_2$}&
\colhead{$\tilde\alpha_{\rm rms}$}& \colhead{$\tilde\eta_{\rm rms}$}&
\colhead{$+\tilde\alphaK$}&
\colhead{$-\tilde\alphaM$}&
\colhead{$\Delta\tilde{t}$}\\
 }
  \startdata
A&     2 &  0.0 &  0.0 &$  0.70 \pm  0.03 $&$  0.67 \pm  0.07 $&$ 1.57 $&$ -0.14 \pm  0.01 $&$  0.04 \pm  0.05 $&$ -0.02 \pm  0.06 $&$  0.09 $&$  0.12 $&$  1.03 $&$  0.01 $&$   150 $\\
B&     4 &  0.9 &  0.4 &$  0.44 \pm  0.01 $&$  0.58 \pm  0.04 $&$ 0.73 $&$  0.00 \pm  0.00 $&$  0.33 \pm  0.02 $&$ -0.11 \pm  0.03 $&$  0.10 $&$  0.21 $&$  1.02 $&$  0.31 $&$   422 $\\
C&    12 &  1.7 &  0.7 &$  0.24 \pm  0.01 $&$  0.46 \pm  0.02 $&$ 0.25 $&$  0.00 \pm  0.00 $&$  0.37 \pm  0.02 $&$ -0.04 \pm  0.01 $&$  0.09 $&$  0.16 $&$  1.00 $&$  0.55 $&$   601 $\\
D&    30 &  1.9 &  0.8 &$  0.16 \pm  0.01 $&$  0.36 \pm  0.02 $&$ 0.11 $&$ -0.00 \pm  0.01 $&$  0.37 \pm  0.02 $&$  0.03 \pm  0.03 $&$  0.07 $&$  0.14 $&$  1.02 $&$  0.62 $&$   350 $\\
E&    60 &  2.0 &  0.8 &$  0.09 \pm  0.01 $&$  0.22 \pm  0.02 $&$ 0.05 $&$  0.00 \pm  0.01 $&$  0.33 \pm  0.01 $&$  0.05 \pm  0.01 $&$  0.09 $&$  0.22 $&$  1.00 $&$  0.66 $&$   711 $\\
F&   150 &  2.0 &  0.9 &$  0.07 \pm  0.00 $&$  0.19 \pm  0.01 $&$ 0.02 $&$  0.01 \pm  0.01 $&$  0.24 \pm  0.05 $&$  0.08 \pm  0.01 $&$  0.07 $&$  0.16 $&$  1.01 $&$  0.69 $&$   225 $\\
G&   300 &  1.8 &  0.9 &$  0.06 \pm  0.00 $&$  0.15 \pm  0.00 $&$ 0.01 $&$  0.01 \pm  0.01 $&$  0.21 \pm  0.02 $&$  0.05 \pm  0.02 $&$  0.06 $&$  0.16 $&$  1.01 $&$  0.66 $&$   177 $\\
H&   600 &  1.8 &  0.9 &$  0.05 \pm  0.01 $&$  0.13 \pm  0.01 $&$ 0.005$&$  0.01 \pm  0.04 $&$  0.14 \pm  0.05 $&$  0.04 \pm  0.01 $&$  0.05 $&$  0.10 $&$  1.03 $&$  0.64 $&$    44 $\\
  \enddata
\end{deluxetable*} 

The consistency of the results for $\alpha$ and $\etat$
with the presence of a steady state
can be assessed by calculating the
growth rate, $\lambda$, of the associated
kinematic mean field dynamo
for a Beltrami field with $k_m=-k_1$,
i.e.\ $\lambda=-\alpha k_1-(\etat+\eta) k_1^2$.
In the saturated state $\lambda$ should vanish.
Again, we present $\lambda$ in nondimensional form,
here in terms of the turbulent decay rate,
\EQ
\tilde\lambda\equiv\lambda/(\eta_{t0}k_1^2)
=\tilde\alpha\tilde\kf
-(\tilde\etat+\tilde\eta). \label{lambda-kin}
\EN
where $\tilde\alpha=\alpha/\alpha_0$ and $\tilde\kf=\kf/k_1$.
Within error bars, the value of $\lambda$ is consistent with zero,
thus supporting the consistency of $\alpha$ and $\etat$ with the
established steady state; see \Tab{Tsum}.
(An exception is Run~A, because it is subcritical and so $\lambda<0$.)
This in turn supports
the applicability of the testfield method to the nonlinear case.
However, as in almost all supercritical runs a small-scale dynamo is operative,
our results which are derived under the assumption of its influence being negligible
may contain a systematic error. If present, it should be small though, given the good precision of the results for $\lambda$.
A more thorough study of the role of the small-scale dynamo will be the subject of future work.

A measure of the reliability of the averages is the length
of the time series in ``turnover'' times,
$\Delta\tilde{t}=u_{\rm rms}k_{\rm f}(t_{\max}-t_{\min})$.
Our results presented in \Tab{Tsum} show a decline of
$\tilde\alpha$ by a factor $\approx15$
and a decline of $\tilde\etat$ by a factor $\approx 5$
as $\Rm$ increases by a factor 300 while $\meanB=0.95...1.4B_{\rm eq}$.

As expected, there are random fluctuations of $\alpha$ and $\etat$,
represented here by their non-dimensional rms values,
$\tilde\alpha_{\rm rms}=\alpha_{\rm rms}/\alpha_0$ and
$\tilde\eta_{\rm rms}=\eta_{\rm rms}/\eta_{t0}$.
Even for large $\Rm$ the fluctuations remain around 0.1 and 0.2, respectively.
This is less than in the kinematic case (Brandenburg et al.\ 2008),
but still comparable to the mean values of $\tilde\alpha$ and $\tilde\etat$,
respectively.

\section{Discussion}

\begin{figure}[t!]
\centering\includegraphics[width=\columnwidth]{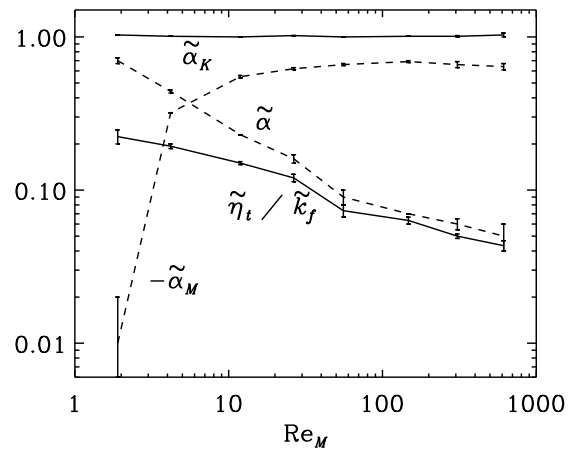}\caption{
$\Rm$-dependence of $\tilde\alpha$ and $\tilde\etat/\tilde\kf$
together with $\tilde\alphaK$ and $-\tilde\alphaM$.
}\label{palpeta}\end{figure}

Let us now put our results in relation to earlier work,
which mostly used mean fields defined as full volume averages,
hence being uniform.
In that case $\alpha$ was quenched all the way to zero
like $\Rm^{-1}$.
This result can be understood in terms of a mutual cancelation
of kinetic and magnetic contributions to the $\alpha$ effect
(Pouquet et al.\ 1976),
\EQ
\alpha=\alphaK+\alphaM,\quad
\alphaK=-\onethird\tau\overline{\oo\cdot\uu},\quad
\alphaM=\onethird\tau\overline{\jj\cdot\bb}/\meanrho,
\label{alphaM}
\EN
where $\oo=\nab\times\uu$.
Assuming $\tau u_{\rm rms}\kf\approx1$
(Brandenburg \& Subramanian 2007), we estimate $\tau$ and hence,
by measuring $\bra{\oo\cdot\uu}$ and $\bra{\jj\cdot\bb}$, we determine
$\tilde\alphaK = \alphaK/\alpha_0$ and $\tilde\alphaM = \alphaM/\alpha_0$;
see \Tab{Tsum} and \Fig{palpeta}.
It turns out that $\tilde\alphaK$ is essentially independent of $\Rm$
[but of course dependent on $\meanB$; see Table~1 of Brandenburg (2001)]
and $\tilde\alphaM$
approaches a certain fraction of $\tilde\alphaK$, reducing the residual
$\alpha$ in \Eq{alphaM} as $\Rm$ increases.
This agrees only qualitatively with the measured decline of $\tilde\alpha$,
because the residual $\alpha$ is sill too big.
However, Eq.~(\ref{alphaM}) assumes isotropy and that the values of $\tau$
are the same for $\alphaK$ and $\alphaM$, which is not borne out by simulations
(Brandenburg \& Subramanian 2007).
By contrast, our direct calculations show that
$\tilde\alpha$ is quenched to values of order
$\tilde\etat/\tilde\kf$, as is necessary for a steady state; see
Eq.~(\ref{lambda-kin}).
Note that the decline of $\tilde\etat$ is
much weaker than in the two-dimensional case
where $\etat$ decreases like $\Rm^{-1}$ (Cattaneo \& Vainshtein 1991).

\section{Conclusions}

For the first time it has been possible to determine both $\alpha_{ij}$
and $\eta_{ij}$ in the magnetically quenched case.
These tensors are here
characterized by the non-tensorial quantities $\alpha$ and $\eta_{\rm t}$.
The consistency of the results of the testfield method suggests that
the nonlinear $\alpha$ can be determined by the knowledge of just $\uu(\xx,t)$
over the past several correlation times--even if it is already influenced by $\meanBB$.
Qualitatively, the quenching of $\alpha$ can be explained by $-\alphaM$
approaching $\alphaK$ for finite field strengths and large $\Rm$.
Generally, $\alpha$ will be quenched to whatever is the value of
$(\etat+\eta)k_1$ (BB02).
However, until now we had no idea how big the quenched value of
$\etat$ is.
There was the possibility that $\etat$ was quenched to very small values,
just like in the two-dimensional case (Cattaneo \& Vainshtein 1991).
If that were true, $\alpha$ would also be very small.
We can now say that this is not the case, because $\etat$ is only reduced
to about 20\% of the kinematic value,
while the normalized value $\tilde\alpha$ is quenched to
$\approx\tilde\etat/\tilde\kf\approx$ 7\% of its kinematic value,
as is seen in \Fig{palpeta}.

Obvious extensions of this work include the application to
non-Beltrami fields and to domains with boundaries and/or shear.
In the latter case there exists a great deal of earlier work with relevant
simulation data supporting the idea of an $\alpha$ effect that is
strongly controlled by magnetic helicity evolution, and that
catastrophic quenching can be decisively alleviated in the presence
of shear-driven magnetic helicity fluxes.

\acknowledgements
We thank Eric Blackman and Alexander Hubbard for useful discussions.
We acknowledge Nordita and the KITP for providing a stimulating atmosphere
during their programs on dynamo theory.
This research was supported in part by the National Science
Foundation under grant PHY05-51164.
We are also grateful for computing resources provided by the
Swedish National Allocations Committee at the
National Supercomputer Centre in Link\"oping.



\begin{thebibliography}{}

\bibitem[]{1036}
Blackman, E. G., \& Brandenburg, A.\yapj{2002}{579}{359}

\bibitem[]{1038}
Brandenburg, A.\yapj{2001}{550}{824}

\bibitem[]{1044}
Brandenburg, A., \& Subramanian, K.\yan{2007}{328}{507}

\bibitem[]{1051}
Brandenburg, A., R\"adler, K.-H., Rheinhardt, M.,
et al.\yapj{2008}{676}{740}

\bibitem[]{1054}
Cattaneo, F., \& Vainshtein, S. I.\yapjl{1991}{376}{L21}

\bibitem[]{1055}
Cattaneo, F., \& Hughes, D. W.\ypre{1996}{54}{R4532}

\bibitem[]{1060}
Kleeorin, N., \& Rogachevskii, I.\ypre{1999}{59}{6724}

\bibitem[]{1061}
Kleeorin, N. I., \& Ruzmaikin, A. A.\yjour{1982}{Magne\-to\-hydro\-dynamics}{18}{116}

\bibitem[]{1082}
Pouquet, A., Frisch, U., \& L\'eorat, J.\yjfm{1976}{77}{321}

\bibitem[]{1094}
Sur, S., Brandenburg, A., \& Subramanian, K.\ymn{2008}{385}{L15}

\bibitem[]{1095}
Vainshtein, S. I., \& Cattaneo, F.\yapj{1992}{393}{165}

\bibitem[]{1100}
Yousef, T. A., Brandenburg, A., \& R\"udiger, G.\yana{2003}{411}{321}

\end{thebibliography}
\end{document}